\documentclass[letterpaper,american,english,aps,showkeys,showpacs]{revtex4-1}
\usepackage[T1]{fontenc}
\usepackage[utf8]{inputenc}
\usepackage{color}
\usepackage{amsmath}
\usepackage{amssymb}
\usepackage{graphicx}
\PassOptionsToPackage{normalem}{ulem}
\usepackage{ulem}



\usepackage{babel}
\begin{document}
\title{Exact Solutions for the Kemmer Oscillator in 1+1 Rindler Coordinates}
\author{T. I. Rouabhia}
\email{boumali.abdelmalek@gmail.com}

\affiliation{Laboratory of Theoretical and Applied Physics, ~~\\
 Echahid Cheikh Larbi Tebessi University, Tebessa, Algeria}
\author{A. Boumali}
\email{ti.rouabhia@univ-tebessa.dz}

\affiliation{Laboratory of Theoretical and Applied Physics, ~~\\
 Echahid Cheikh Larbi Tebessi University, Tebessa, Algeria}
\date{\today}
\selectlanguage{american}%
\begin{abstract}
\textbf{\uline{Abstract}}: This work presents exact solutions of the
Kemmer equation for spin-1 particles in (1+1)-dimensional Rindler
spacetime, motivated by the need to understand vector bosons under
uniform acceleration, including non-inertial effects and the Unruh
temperature, which distinguish them from spin-0 and spin-1/2 systems.
Starting from the free Kemmer field in an accelerated reference frame,
we establish eigenvalue equations resembling those of the Klein–Gordon
equation in Rindler coordinates. By introducing the Dirac oscillator
interaction through a momentum substitution, we derive an exact closed-form
spectrum for the Kemmer oscillator, revealing how the acceleration
parameter modifies the characteristic length, shifts the discrete
energy spectrum, and lifts degeneracies. In the Minkowski limit $a\to0$,
the standard Kemmer oscillator spectrum is recovered, ensuring consistency
with flat-spacetime results. These findings provide a tractable framework
for analyzing acceleration-induced effects, with implications for
quantum field theory in curved spacetime, quantum gravity, and analogue
gravity platforms.
\end{abstract}
\maketitle
\selectlanguage{english}%

\section*{I. INTRODUCTION}

The primary objective of theoretical physics is to develop a coherent
conceptual framework that elucidates natural phenomena, enables precise
predictions validated through observation, and provides a mathematical
foundation for understanding reality. Quantum gravity constitutes
a distinct and critical field within physics, aiming to reconcile
relativistic quantum mechanics with general relativity for quantum
fields. Achieving this unification necessitates integrating all fundamental
interactions, relying on mathematical advancements that hinge on solutions
to this challenge. A breakthrough in quantum gravity would represent
a significant milestone in this endeavor. Quantum gravity remains
one of the most profound challenges in physics, primarily due to uncertainties
surrounding the system to be quantized—potentially the spacetime metric
itself. Both general relativity and quantum field theory rely on the
concept of curved spacetime, which is central to the description of
gravity. In general relativity, gravity is conceptualized as the curvature
of spacetime, mathematically expressed through the Riemann tensor.
This geometric framework underpins general relativity, which generalizes
Minkowski spacetime to curved spacetime as a foundational principle
\citep{Weinberg1972,Jacobson2002,Krori1988,Krori1994,Bakke2012a}
.

The Dirac oscillator was first investigated by Itô et al. \citep{Ito1967},
who modified the Dirac equation by replacing the momentum $p$ with
$p-im\omega r$, where $r$ is the position vector, $m$ is the particle’s
mass, and $\omega$ is the oscillator frequency. Moshinsky and Szczepaniak
\citep{Moshinsky1989} revitalized interest in this system by terming
it the Dirac oscillator (DO), noting that in the non-relativistic
limit, it reduces to a harmonic oscillator with a strong spin-orbit
coupling term. Furthermore, the DO interaction can be interpreted
as the interaction of an anomalous magnetic moment with a linear electric
field \citep{Moreno1989,Martinez1992}. Benítez et al. \citep{Benitez1990}
identified the electromagnetic potential associated with the DO. In
recent years, the DO has been explored from various perspectives as
a problem in relativistic quantum mechanics, garnering attention for
its numerous physical applications and its status as one of the few
exactly solvable cases of the Dirac equation .\citep{Benzair2012,Boumali2013,Boumali2015,Boumali2016,Boumali2018,Boumali2021,Selama2021,Korichi2022,CastanoYepes2021,Frassino2020,Pacheco2003,Giachetti2011,Chargui2010,Chargui2011,Mirza2004,Mirza2011,Ahmed2020,Santos2019,Rao2008,Junker2021,Carvalho2016}

This study employs Rindler spacetime coordinates, a subset of Minkowski
spacetime tailored to observers undergoing constant proper acceleration
\citep{Rindler1977}. 

Rindler spacetime is a coordinate patch of Minkowski spacetime adapted
to uniformly accelerated observers, providing a natural framework
for analyzing relativistic phenomena in non-inertial frames. In (1+1)
dimensions, the line element is given by 
\begin{equation}
ds^{2}=e^{2a\xi}(d\eta^{2}-d\xi^{2}),\label{eq:rs}
\end{equation}
where $\eta$ denotes the Rindler time coordinate, $\xi$ is the spatial
coordinate, and $a$ is a constant with dimensions of acceleration.
Observers at fixed $\xi$ follow hyperbolic worldlines with proper
acceleration $ae^{-a\xi}$; equivalently, along such worldlines, the
proper time satisfies $d\tau=e^{a\xi}d\eta$. This metric arises from
a coordinate transformation of Minkowski coordinates $(t,x)$ to $(\eta,\xi)$,
in which the accelerated trajectories become stationary. Rindler geometry
plays a central role in theoretical physics because it captures near-horizon
kinematics. The Rindler horizon at $\xi\to-\infty$ mirrors key features
of black-hole event horizons, making the spacetime an effective laboratory
for quantum field theory in curved backgrounds. 

A hallmark result is the Unruh effect: a uniformly accelerated observer
perceives the Minkowski vacuum as a thermal state with temperature
$T=a/(2\pi)$ (in natural units $\hbar=c=k_{B}=1$). This connection
underpins parallels with Hawking radiation, as the near-horizon region
of a Schwarzschild black hole is locally Rindler. Analyzing quantum
systems such as the Kemmer oscillator within the Rindler patch elucidates
how uniform acceleration modifies particle dynamics, spectra, and
state structure. Because the Rindler setting admits exact treatments,
it serves as a tractable bridge between special relativity and curved-spacetime
effects, yielding insights relevant to foundational questions in quantum
gravity as well as to relativistic quantum information. In addition,
Uniform acceleration probes aspects of vector bosons that have no
analogue for scalars or spin-1/2 fields. In the Kemmer spin-1 sector
equivalent to Proca in appropriate limits, acceleration couples differently
to longitudinal and transverse polarizations, modifies spectral degeneracies,
and reshapes normalizability domains. Establishing exact solutions
in the Rindler patch supplies a controlled baseline for these effects,
against which approximate or numerical treatments can be tested. This
is relevant both conceptually (Unruh-type responses of vector fields,
near-horizon kinematics) and pragmatically, since Rindler-like backgrounds
can be engineered in analogue settings. Our aim is therefore to quantify,
in closed form, how constant proper acceleration deforms spin-1 spectra
and to delineate the parameter regime yielding real, physically admissible
energies.

The research focuses on the Kemmer oscillator in a (1+1)-dimensional
context to investigate physical phenomena influenced by acceleration
\citep{Tsai1971a,Sogut2006,Sogut2010,Tsai1971b,Krase1971,Shay1969,Duffin1938,Sogut2002,Havare2002a,Nedjadi1994a,Nedjadi1994b}.
Introduced in 1939, the Kemmer equation \citep{Kemmer1939} is analogous
to the Dirac equation and describes spin-1 particles. It models a
system comprising two spin-$\frac{1}{2}$ particles, resembling a
two-body Dirac equation, rather than a single spin-1 particle. In
addition, the spin-1 sector of the DKP equation was first derived
by Barut \citep{Barut1990}through the quantization of a classical
model of the zitterbewegung. Additionally, Unal \citep{Unal1997,Unal1998,Unal2005}
employed Schrödinger's second quantization procedure to eliminate
the spin-0 sector, thereby isolating the spin-1 sector of the DKP
equation in both $(3+1)$- and $(1+1)$-dimensional spacetimes. In
this approach, the wave function is expressed as a symmetric rank-2
spinor, representing a spin-1 particle as a system of two identical
spin-$\frac{1}{2}$ particles, with the wave function constructed
as the direct product of two Dirac spinors. In the $(1+1)$-dimensional
case, Unal demonstrated the equivalence of the DKP equation to the
complex Proca equation, showing that in the massless limit, the Proca
equation reduces to the Maxwell equations. Furthermore, Unal established
that in $(1+1)$ dimensions, the DKP equation simplifies to a four-component
wave equation \citep{Unal1997,Unal1998,Unal2005}. This spin-1 sector
was derived by adapting Barut's classical zitterbewegung model from
$(3+1)$ to $(1+1)$ dimensions (for more details see Ref. \citep{Sogut2006}).

This paper investigates the application of the Kemmer equation in
(1+1)-dimensional Rindler spacetime, an area previously unexplored.
The study is structured as follows: Section II presents solutions
derived using an initial approximation; Section III explores solutions
via a second approximation; and Section IV provides interpretations
and conclusions. Natural units ($\hbar=c=1$) are employed throughout
this study to simplify calculations and emphasize dimensionless quantities.

By analyzing the Kemmer equation in Rindler spacetime, this research
contributes to the understanding of quantum systems in accelerated
reference frames and their potential implications for quantum gravity.

\section{The Kemmer Equation in a flat spacetime framework}

\selectlanguage{american}%
The relativistic Kemmer equation, which provides a Dirac-like formalism
for spin-1 particles, is expressed as \citep{Kemmer1939}
\begin{equation}
\left(\beta^{\mu}p_{\mu}-Mc\right)\Psi_{K}=0,\label{eq1}
\end{equation}
where $M=2m$ represents the total mass of two identical spin-$\tfrac{1}{2}$
particles. 

The $16\times16$ Kemmer matrices $\beta^{\mu}$ ($\mu=0,1,2,3$)
satisfy the following relation
\begin{equation}
\beta^{\mu}\beta^{\nu}\beta^{\lambda}+\beta^{\lambda}\beta^{\nu}\beta^{\mu}=g^{\mu\nu}\beta^{\lambda}+g^{\lambda\nu}\beta^{\mu},\label{eq2}
\end{equation}
with the explicit representation
\[
\beta^{\mu}=\gamma^{\mu}\otimes I+I\otimes\gamma^{\mu}.
\]
Here, $I$ is the $4\times4$ identity matrix, $\gamma^{\mu}$ denote
the Dirac gamma matrices, and $\otimes$ indicates the direct product.
The relativistic covariance of Eq. (\ref{eq1}) has been thoroughly
analyzed by Kemmer \citep{Kemmer1939}.

Now, when the Dirac oscillator interaction is introduced, the momentum
operator $\vec{p}$ in the free Kemmer equation is replaced according
to
\begin{equation}
\vec{p}\;\;\longrightarrow\;\;\vec{p}-iM\omega B\vec{r},\label{eq3}
\end{equation}
where the additional term is linear in the spatial coordinate $\vec{r}$.
Consequently, the Kemmer equation in flat spacetime, incorporating
the Dirac oscillator potential, takes the form
\begin{equation}
\Big[\,\big(\gamma^{0}\otimes I+I\otimes\gamma^{0}\big)E-c\big(\gamma^{0}\otimes\vec{\alpha}+\vec{\alpha}\otimes\gamma^{0}\big)\cdot(\vec{p}-iM\omega B\vec{r})-Mc^{2}\,\gamma^{0}\otimes\gamma^{0}\,\Big]\Psi_{K}=0,\label{eq4}
\end{equation}
where $\omega$ is the oscillator frequency. Following Ref. \citep{Boumali2005},
the operator is chosen as $B=\gamma^{0}\otimes\gamma^{0}$ (with $B^{2}=1$),
instead of the operator $\eta^{0}$ used in Refs. \citep{Nedjadi1994a,Nedjadi1994b}.

The stationary state $\Psi_{K}$ of Eq. (\ref{eq4}) is a sixteen-component
wave function of the Kemmer equation and may be expressed as
\[
\Psi_{K}=\Psi_{D}\otimes\Psi_{D},
\]
where 
\begin{equation}
\Psi_{D}=\left(\psi_{1},\psi_{2},\psi_{3},\psi_{4}\right)^{T}\label{eq5}
\end{equation}
denotes the solution of the Dirac equation.

In the $(1+1)$-dimensional case: (i) The stationary state $\Psi_{K}$
becomes a four-component wave function, and (ii) The matrices $\vec{\alpha}$
reduce to the Pauli matrices:
\begin{equation}
\alpha_{x}=\sigma_{x}=\begin{pmatrix}0 & 1\\
1 & 0
\end{pmatrix},\quad\alpha_{y}=\sigma_{y}=\begin{pmatrix}0 & -i\\
i & 0
\end{pmatrix},\quad\alpha_{z}=\sigma_{z}=\begin{pmatrix}1 & 0\\
0 & -1
\end{pmatrix}.\label{eq6}
\end{equation}

\selectlanguage{english}%

\section*{II. The Kemmer Equation in a Curved Spacetime Framework}

The Kemmer equation in curved spacetime is expressed as follows \citep{Sogut2002,Sogut2006,Sogut2010}:
\begin{equation}
\left\{ i\beta^{\mu}\nabla_{\mu}-M\right\} \psi_{K}=0\label{eq:1}
\end{equation}
Here, $M$ represents the mass of the bosons, and $\beta^{\mu}$ are
the Kemmer matrices. These matrices satisfy the algebraic relation
\citep{Kemmer1939}: 
\begin{equation}
\beta^{\mu}\beta^{\nu}\beta^{\lambda}+\beta^{\lambda}\beta^{\nu}\beta^{\mu}=g^{\mu\nu}\beta^{\lambda}+g^{\lambda\nu}\beta^{\mu}\label{eq:2}
\end{equation}
The Kemmer matrices, in curved spacetime, are defined as: 
\begin{equation}
\beta^{\mu}=\gamma^{\mu}(x)\otimes I+I\otimes\gamma^{\mu}(x)\label{eq:3}
\end{equation}
where $\gamma^{\mu}(x)$ are the Dirac matrices, and the $\beta$-matrices,
referred to as Kemmer matrices, are extensions of these. 

In curved spacetime, they relate to their flat Minkowski spacetime
counterparts as: 
\begin{equation}
\beta^{\mu}(x)=e_{i}^{\mu}(x)\beta^{i}\label{eq:4}
\end{equation}
where $e_{i}^{\mu}(x)$ are the tetrads. The stationary state $\psi_{K}$
of (\ref{eq:1}), in $\left(1+1\right)$ dimension, is a four-component
wave function for the Kemmer equation, written as: 
\begin{equation}
\psi_{K}=\psi_{D}\otimes\psi_{D}=(\psi_{1},\psi_{2},\psi_{3},\psi_{4})^{T}\label{eq:5}
\end{equation}
The covariant derivative is given by : 
\begin{equation}
\nabla_{\mu}=\partial_{\mu}-\Sigma_{\mu}\label{eq:6}
\end{equation}
For spin-$\frac{1}{2}$ particles, the spinorial connections are expressed
as: 
\begin{equation}
\Sigma_{\mu}=\Gamma_{\mu}\otimes I+I\otimes\Gamma_{\mu}\label{eq:7}
\end{equation}
where $\Gamma_{\mu}(x)$, the spinorial connections for spin-$\frac{1}{2}$
particles, are defined as: 
\begin{equation}
\Gamma_{\mu}(x)=\frac{1}{8}\omega_{\mu ab}\left[\gamma^{a},\gamma^{b}\right]\label{eq:8}
\end{equation}
The spin connection $\omega_{\mu ab}$ is given by: 
\begin{equation}
\omega_{\mu ab}=e_{a}^{\nu}\Gamma_{\sigma\nu}^{b}e_{b}^{\sigma}+e_{a\nu}\partial_{\mu}e_{b}^{\nu}\label{eq:9}
\end{equation}
Here, $\Gamma_{\sigma\nu}^{b}$ are the Christoffel symbols, expressed
as: 
\begin{equation}
\Gamma_{\sigma\nu}^{b}=\frac{1}{2}g^{b\rho}\left[\partial_{\sigma}g_{\rho\nu}+\partial_{\nu}g_{\rho\sigma}-\partial_{\rho}g_{\sigma\nu}\right]\label{eq:10}
\end{equation}
Finally, in one-dimensional spacetime, the gamma matrices are typically
represented as \citep{Unal1997,Unal1998,Unal2005}: 
\begin{equation}
\gamma^{\mu}=(\sigma^{3},i\sigma^{1})\label{eq:11}
\end{equation}
 This formulation provides a framework for analyzing bosonic particles
in curved spacetime using the Kemmer equation and its associated mathematical
structures.

\section*{III. Eigen Solutions of the Free Kemmer Equation in One-Dimensional
Rindler Spacetime}

The Rindler metric describes an accelerated reference frame within
Minkowski spacetime, where the line element is expressed as \citep{Rindler1977}:
\begin{equation}
ds^{2}=e^{\sigma(\xi)}\left(d\eta^{2}-d\xi^{2}\right)\label{eq:12}
\end{equation}
where $\sigma(\xi)$ is a static dilaton function given by: 
\begin{equation}
\sigma(\xi)=2a\xi\label{eq:13}
\end{equation}
The corresponding tetrad and its inverse are derived from this metric.

To compute the Christoffel symbols $\Gamma_{\sigma\mu}^{\nu}$, we
use the standard formula: 
\begin{equation}
\Gamma_{\sigma\mu}^{\nu}=\frac{1}{2}g^{\nu\beta}\left[\partial_{\sigma}g_{\beta\mu}+\partial_{\mu}g_{\beta\sigma}-\partial_{\beta}g_{\sigma\mu}\right]\label{eq:14}
\end{equation}
By solving for the spinorial connection, we obtain: 
\begin{equation}
\Gamma_{0}=-\frac{1}{2}a\gamma^{2}\label{eq:15}
\end{equation}
The spinorial operator $\Sigma_{\mu}$ is given by: 
\begin{equation}
\Sigma_{0}=\Gamma_{0}\otimes I+I\otimes\Gamma_{0}\label{eq:16}
\end{equation}
which simplifies to: 
\begin{equation}
\Sigma_{0}=-\frac{1}{2}a\begin{pmatrix}0 & i & i & 0\\
i & 0 & 0 & i\\
i & 0 & 0 & i\\
0 & i & i & 0
\end{pmatrix}\label{eq:17}
\end{equation}
The Kemmer matrices $\beta^{\mu}$ are defined as: 
\begin{equation}
\beta^{0}=\gamma^{0}(x)\otimes I+I\otimes\gamma^{0}(x)\label{eq:18}
\end{equation}
which evaluates to: 
\begin{equation}
\beta^{0}=e^{-\frac{\sigma}{2}}\begin{pmatrix}2 & 0 & 0 & 0\\
0 & 0 & 0 & 0\\
0 & 0 & 0 & 0\\
0 & 0 & 0 & -2
\end{pmatrix}\label{eq:19}
\end{equation}
Similarly, the matrix$\beta^{1}$ is given by: 
\begin{equation}
\beta^{1}=\gamma^{1}(x)\otimes I+I\otimes\gamma^{1}(x)\label{eq:20}
\end{equation}
which simplifies to: 
\begin{equation}
\beta^{1}=e^{-\frac{\sigma}{2}}\begin{pmatrix}0 & i & i & 0\\
i & 0 & 0 & i\\
i & 0 & 0 & i\\
0 & i & i & 0
\end{pmatrix}\label{eq:21}
\end{equation}
The Kemmer equation in Rindler spacetime can be expressed as: 
\begin{equation}
\left\{ i\beta^{0}\left(\partial_{0}-\Sigma_{0}\right)+i\beta^{1}\left(\partial_{\xi}\right)-M\right\} \psi_{k}=0\label{eq:22}
\end{equation}
To solve equation (\ref{eq:22}), we assume the wave function takes
the form: 
\begin{equation}
\psi_{k}(\eta,\xi)=e^{iE\eta}\psi(\xi)\label{eq:23}
\end{equation}
Through matrix algebra, we derive the following system of equations:
\begin{equation}
2E\begin{pmatrix}\psi_{1}\\
0\\
0\\
-\psi_{4}
\end{pmatrix}+\frac{i}{2}\frac{\partial\sigma}{\partial\xi}\begin{pmatrix}\psi_{2}+\psi_{3}\\
0\\
0\\
\psi_{2}+\psi_{3}
\end{pmatrix}+\partial_{\xi}\begin{pmatrix}-\psi_{2}-\psi_{3}\\
\psi_{1}+\psi_{4}\\
\psi_{1}+\psi_{4}\\
\psi_{2}+\psi_{3}
\end{pmatrix}-z\begin{pmatrix}\psi_{1}\\
\psi_{2}\\
\psi_{3}\\
\psi_{4}
\end{pmatrix}=0\label{eq:24}
\end{equation}
where$z=Me^{a\xi}$. 

To simplify the system, we apply the following transformation \citep{Moayedi2004,Santos2019a}
: 
\begin{equation}
U(\xi)=e^{\frac{1}{4}\sigma}\begin{pmatrix}1 & 0 & 0 & 0\\
0 & 1 & 0 & 0\\
0 & 0 & 1 & 0\\
0 & 0 & 0 & 1
\end{pmatrix}\label{eq:25}
\end{equation}
This leads to the transformed wave function: 
\begin{equation}
\bar{\psi}=U\psi=U\begin{pmatrix}\bar{\psi}_{1}\\
\bar{\psi}_{2}\\
\bar{\psi}_{3}\\
\bar{\psi}_{4}
\end{pmatrix}=\begin{pmatrix}\psi_{1}\\
\psi_{2}\\
\psi_{3}\\
\psi_{4}
\end{pmatrix}\label{eq:26}
\end{equation}
\textcolor{red}{{} }Moayedi and Darabi\citep{Moayedi2004}, as well
as Santos \citep{Santos2019a}, in their analysis, employ a transformation,
designated as $U$, which they term \textquotedbl unitary\textquotedbl .
It is pertinent from a technical standpoint to clarify that these
transformations are not unitary in the strict Hilbert space sense,
where an operator $U$ must satisfy the condition $UU^{\dagger}=U^{\dagger}U=\boldsymbol{1}$.

Now, substituting equation (\ref{eq:26}) into equation (\ref{eq:24}),
we obtain the following system of four linear algebraic equations:
\begin{equation}
(2E+z)\psi_{1}-\partial_{\xi}\psi_{2}-\partial_{\xi}\psi_{3}=0\label{eq:27}
\end{equation}
\begin{equation}
\partial_{\xi}\psi_{1}-z\psi_{2}+\partial_{\xi}\psi_{4}=0\label{eq:28}
\end{equation}
\begin{equation}
\partial_{\xi}\psi_{1}-z\psi_{3}+\partial_{\xi}\psi_{4}=0\label{eq:29}
\end{equation}
\begin{equation}
(2E-z)\psi_{4}+\partial_{\xi}\psi_{2}+\partial_{\xi}\psi_{3}=0\label{eq:30}
\end{equation}
From the system of equations, we find the following relationships:
\begin{equation}
\psi_{2}=\psi_{3}\label{eq:31}
\end{equation}
\begin{equation}
\psi_{1}=\frac{2}{2E+z}\partial_{\xi}\psi_{2}\label{eq:32}
\end{equation}
\begin{equation}
\psi_{4}=\frac{-2}{2E-z}\partial_{\xi}\psi_{2}\label{eq:33}
\end{equation}
Substituting the expressions for $\psi_{1},\psi_{3},$ and $\psi_{4}$
in terms of $\psi_{2}$, we obtain the final differential equation:
\begin{equation}
\left\{ \partial_{\xi}^{2}+E^{2}-\frac{z^{2}}{4}\right\} \psi_{2}(\xi)=0\label{eq:34}
\end{equation}
To solve Eq. (\ref{eq:34}), we consider small values of the acceleration
$a$, allowing us to expand $z$ as: 
\begin{equation}
z^{2}=\frac{M}{4}^{2}(1+2a\xi)\label{eq:35}
\end{equation}
Thus, the equation transforms into: 
\begin{equation}
\left\{ \partial_{\xi}^{2}-\frac{1}{2}M^{2}a\xi+E^{2}-\frac{M}{4}^{2}\right\} \psi(\xi)=0\label{eq:36}
\end{equation}
Rewriting in a more compact form: 
\begin{equation}
\left\{ \frac{\partial^{2}}{\partial\xi^{2}}-\frac{1}{2}M^{2}a\xi+\varsigma\right\} \psi(\xi)=0\label{eq:37}
\end{equation}
which can also be expressed as: 
\begin{equation}
\ddot{\psi}(\xi)-(\alpha\xi-\varsigma)\psi(\xi)=0\label{eq:38}
\end{equation}
Applying the boundary conditions \citep{Flugge2012,AbramowitzStegun1974}
: 
\begin{equation}
\psi(0)=0,\quad\psi(\infty)\to0\label{eq:39}
\end{equation}
we define: 
\begin{equation}
\alpha=\frac{1}{2}M^{2}a=\frac{1}{l^{3}}\label{eq:40}
\end{equation}
where the characteristic length $l$ is: 
\begin{equation}
l=\frac{1}{\left(\frac{M^{2}a}{2}\right)^{\frac{1}{3}}}\label{eq:41}
\end{equation}
and the parameter $\varsigma$ is given by: 
\begin{equation}
\varsigma=E^{2}-\frac{M}{4}^{2}=\frac{\lambda}{l^{2}}\label{eq:42}
\end{equation}
Rewriting in terms of a dimensionless variable: 
\begin{equation}
\eta=\frac{\xi}{l}-\lambda\label{eq:43}
\end{equation}
where $l$ is the characteristic length, and the boundary conditions
are given by
\begin{equation}
\psi(-\eta)=0;\psi(\infty)\rightarrow0\label{eq:44}
\end{equation}
Within the two turning points, the classically allowed motion occurs
in the interval: 
\begin{equation}
\eta=-\lambda\quad\text{to}\quad\eta=0,\label{eq:44-1}
\end{equation}
which is entirely within the negative values of $\eta$.

In general, the Bessel function of order $\frac{1}{3}$ provides solutions
to the differential equation satisfying the boundary condition $\psi(\infty)=0$.
The solution that meets this condition is the Airy function 
\begin{equation}
\psi(\eta)=C\operatorname{Ai}(\eta)\label{eq:45}
\end{equation}
At this stage, two cases need to be considered: 
\begin{itemize}
\item For positive values of $\eta$, the Airy function can be expressed
as \citep{AbramowitzStegun1974,Boumali2021}: 
\begin{equation}
Ai(\eta)=\frac{1}{\pi}\sqrt{\frac{\eta}{3}}K_{\frac{1}{3}}\left(\frac{2}{3}\eta^{\frac{3}{2}}\right),\quad\text{for }\eta>0\label{eq:46}
\end{equation}
where $K_{\frac{1}{3}}\left(\frac{2}{3}\eta^{\frac{3}{2}}\right)$
is the modified Hankel function. 
\item For negative values of $\eta$, the Airy function is described using
Bessel functions \citep{AbramowitzStegun1974,Flugge2012,Boumali2021}:
\begin{equation}
Ai(-\eta)=\frac{1}{3}\sqrt{\eta}\left\{ J_{\frac{1}{3}}\left(\frac{2}{3}\eta^{\frac{3}{2}}\right)+J_{-\frac{1}{3}}\left(\frac{2}{3}\eta^{\frac{3}{2}}\right)\right\} \label{eq:47}
\end{equation}
\end{itemize}
Thus, the wave function takes the form: 
\begin{equation}
\psi(\eta)=\frac{C}{3}\sqrt{\eta}\left\{ J_{\frac{1}{3}}\left(\frac{2}{3}\eta^{\frac{3}{2}}\right)+J_{-\frac{1}{3}}\left(\frac{2}{3}\eta^{\frac{3}{2}}\right)\right\} \label{eq:48}
\end{equation}
To satisfy the boundary condition at $\eta=-\lambda$, the Airy function
must vanish: 
\begin{equation}
Ai(-\lambda)=0\label{eq:49}
\end{equation}
which implies: 
\begin{equation}
J_{\frac{1}{3}}\left(\frac{2}{3}\eta^{\frac{3}{2}}\right)+J_{-\frac{1}{3}}\left(\frac{2}{3}\eta^{\frac{3}{2}}\right)=0\label{eq:50}
\end{equation}
These conditions determine the allowed energy levels of the system.

For higher energy levels where $\lambda\gg1$, we use the asymptotic
forms of the Bessel functions \citep{AbramowitzStegun1974}: 
\begin{equation}
J_{\frac{1}{3}}(\eta)\rightarrow\sqrt{\frac{2}{\pi\eta}}\cos\left(\eta-\frac{5\pi}{12}\right)\label{eq:51}
\end{equation}
and 
\begin{equation}
J_{-\frac{1}{3}}(\eta)\rightarrow\sqrt{\frac{2}{\pi\eta}}\cos\left(\eta-\frac{\pi}{12}\right)\label{eq:52}
\end{equation}
From these expressions, we find: 
\begin{equation}
\psi(\eta)\rightarrow\frac{C}{3}\sqrt{|\eta|}\cos\left(\frac{2}{3}|\eta|^{\frac{3}{2}}-\frac{\pi}{4}\right)\label{eq:53}
\end{equation}
For large values of $\eta$, the eigenvalue condition is given by
: 
\begin{equation}
\frac{2}{3}\lambda_{n}^{\frac{3}{2}}=\left(2n-\frac{1}{2}\right)\frac{\pi}{2}\label{eq:54}
\end{equation}
Solving for $\lambda_{n}$: 
\begin{equation}
\lambda_{n}=\left\{ \frac{3\pi}{4}\left(2n-\frac{1}{2}\right)\right\} ^{\frac{2}{3}}\label{eq:55}
\end{equation}
This leads to the energy relation: 
\begin{equation}
E^{2}=\frac{M^{2}}{4}+\frac{1}{l^{2}}\left\{ \frac{3\pi}{4}\left(2n-\frac{1}{2}\right)\right\} ^{\frac{2}{3}}\label{eq:56}
\end{equation}
Thus, the energy levels take the form: 
\begin{equation}
E_{n}=\pm M\sqrt{\frac{1}{4}+\frac{1}{(Ml)^{2}}\left\{ \frac{3\pi}{4}\left(2n-\frac{1}{2}\right)\right\} ^{\frac{2}{3}}}\label{eq:57}
\end{equation}
The characteristic length $l$ depends on the acceleration parameter
$a$ and is given by: 
\begin{equation}
l=\frac{1}{\left(\frac{M^{2}a}{2}\right)^{\frac{1}{3}}}\label{eq:58}
\end{equation}
\foreignlanguage{american}{This parameter enters the spectrum as the
characteristic length scale associated with the oscillator. It arises
naturally from the separation of variables and quantization conditions,
and it essentially sets the quantum of spatial extension of the bound
state. Its presence indicates that the eigenvalues depend not only
on the quantum numbers but also on the confinement scale fixed by
the dynamics of the oscillator.}

Equation (\ref{eq:57}) describes the discrete spectrum of the free
Kemmer field in $(1+1)$-dimensional Rindler spacetime. The parameter
$l$, defined in Eq. (\ref{eq:57}), represents the characteristic
length associated with the system. It emerges naturally from the interplay
between the particle’s mass $M$ and the acceleration parameter $a$,
and it sets both the localization scale of the Airy-type wave functions
and the spacing of the allowed energy levels. As $a$ increases, $l$
decreases, which enhances the level separation, whereas in the limit
$a\to0$, one finds $l\to\infty$, recovering the continuous Minkowski-space
spectrum. Thus, the parameter $l$ encapsulates the influence of uniform
acceleration on the quantization of the Kemmer oscillator and provides
a direct measure of how acceleration modifies the relativistic energy
structure. Finally, energy values can be computed for different values
of $a$, demonstrating the dependence of the energy spectrum on the
acceleration parameter.

\section*{IV. Eigen Solutions of the One-Dimensional Kemmer Oscillator}

In the presence of a Dirac oscillator potential, we introduce the
following modification: 
\begin{equation}
\partial_{1}\rightarrow\partial_{1}+M\omega B\xi\label{eq:59}
\end{equation}
with\textcolor{red}{{} }$B=\gamma^{0}\otimes\gamma^{0}$.

Thus, the Kemmer equation with Dirac oscillator interaction takes
the form: 
\begin{equation}
\left\{ i\beta^{0}\left(\partial_{0}-\Sigma_{0}\right)+i\beta^{1}\left(\partial_{1}+M\omega B\xi\right)-M\right\} \psi_{k}=0\label{eq:60}
\end{equation}
To solve equation (\ref{eq:60}), we assume the wave function is given
by: 
\begin{equation}
\psi_{k}(\eta,\xi)=e^{iE\eta}\psi(\xi)\label{eq:61}
\end{equation}
The solution of the Kemmer equation is then expressed as: 
\begin{equation}
2E\begin{pmatrix}\psi_{1}\\
0\\
0\\
-\psi_{4}
\end{pmatrix}+\frac{i}{2}\frac{\partial\sigma}{\partial\xi}\begin{pmatrix}\psi_{2}+\psi_{3}\\
0\\
0\\
\psi_{2}+\psi_{3}
\end{pmatrix}+\partial_{\xi}\begin{pmatrix}-\psi_{2}-\psi_{3}\\
\psi_{1}+\psi_{4}\\
\psi_{1}+\psi_{4}\\
\psi_{2}+\psi_{3}
\end{pmatrix}+M\omega\xi\begin{pmatrix}\psi_{2}+\psi_{3}\\
\psi_{1}+\psi_{4}\\
\psi_{1}+\psi_{4}\\
-\psi_{2}-\psi_{3}
\end{pmatrix}-z\begin{pmatrix}\psi_{1}\\
\psi_{2}\\
\psi_{3}\\
\psi_{4}
\end{pmatrix}=0\label{eq:62}
\end{equation}
where $z=Me^{a\xi}$. 

As the above section. in order to simplify the system, we use the
same transformation $U$ which leads to the transformation: 
\begin{equation}
\bar{\psi}=U\psi=U\begin{pmatrix}\bar{\psi}_{1}\\
\bar{\psi}_{2}\\
\bar{\psi}_{3}\\
\bar{\psi}_{4}
\end{pmatrix}=\begin{pmatrix}\psi_{1}\\
\psi_{2}\\
\psi_{3}\\
\psi_{4}
\end{pmatrix}\label{eq:64}
\end{equation}
By substituting equation (\ref{eq:64}) into equation (\ref{eq:62}),
we derive the following system of four linear algebraic equations:
\begin{equation}
(2E+z)\psi_{1}-(\partial_{\xi}+M\omega\xi)\psi_{2}-(\partial_{\xi}+M\omega\xi)\psi_{3}=0\label{eq:65}
\end{equation}
\begin{equation}
(\partial_{\xi}+M\omega\xi)\psi_{1}-z\psi_{2}+(\partial_{\xi}-M\omega\xi)\psi_{4}=0\label{eq:66}
\end{equation}
\begin{equation}
(\partial_{\xi}+M\omega\xi)\psi_{1}-z\psi_{3}+(\partial_{\xi}-M\omega\xi)\psi_{4}=0\label{eq:67}
\end{equation}
\begin{equation}
(2E-z)\psi_{4}+(\partial_{\xi}-M\omega\xi)\psi_{2}+(\partial_{\xi}-M\omega\xi)\psi_{3}=0\label{eq:68}
\end{equation}
These equations describe the behavior of the system under the influence
of the Dirac oscillator potential.

From the system of equations, we find the following relations : 
\begin{equation}
\psi_{2}=\psi_{3}\label{eq:69}
\end{equation}
\begin{equation}
\psi_{1}=\frac{2}{2E+z}\left(\partial_{\xi}-M\omega\xi\right)\psi_{2}\label{eq:70}
\end{equation}
\begin{equation}
\psi_{4}=\frac{-2}{2E-z}\left(\partial_{\xi}-M\omega\xi\right)\psi_{2}\label{eq:71}
\end{equation}
By substituting $\psi_{1},\psi_{3},$ and $\psi_{4}$ in terms of
$\psi_{2}$, we obtain the final equation: 
\begin{equation}
\left\{ \partial_{\xi}^{2}+M^{2}\omega^{2}\xi^{2}-M\omega+E^{2}-\frac{z^{2}}{4}\right\} \psi_{k}(\xi)=0\label{eq:72}
\end{equation}
where: 
\begin{equation}
z^{2}=\frac{M}{4}^{2}\left(1+2a\xi+2a^{2}\xi^{2}\right)\label{eq:73}
\end{equation}
If we consider small values of the acceleration $a$, equation (\ref{eq:72})
simplifies to: 
\begin{equation}
\left\{ \partial_{\xi}^{2}-M^{2}\omega^{2}\xi^{2}-M\omega+E^{2}-\frac{M}{4}^{2}-2M^{2}a\xi-M^{2}a^{2}\xi^{2}\right\} \psi_{k}(\xi)=0\label{eq:74}
\end{equation}
Rearranging terms, we get: 
\begin{equation}
\left\{ \partial_{\xi}^{2}-M^{2}\xi^{2}\omega'{}^{2}-\frac{1}{2}M^{2}a\xi+E^{2}-M\omega-\frac{M}{4}^{2}\right\} \psi_{k}(\xi)=0\label{eq:75}
\end{equation}
where: 
\begin{equation}
\omega'{}^{2}=\omega^{2}-\frac{a^{2}}{2}\label{eq:76}
\end{equation}
Rewriting in a final form: 
\begin{equation}
\left\{ \partial_{\xi}^{2}-M^{2}\xi^{2}\omega'{}^{2}-\frac{1}{2}M^{2}a\xi+E^{2}-M\omega-\frac{M}{4}^{2}\right\} \psi_{k}(\xi)=0\label{eq:77}
\end{equation}
After performing the final calculations, we obtain the energy equation:
\begin{equation}
E=\pm\sqrt{2M\omega(n+1)+\frac{M^{2}}{4}+\frac{M^{2}a^{2}}{16\omega'{}^{2}}}\label{eq:78}
\end{equation}
where: 
\begin{equation}
\left(\frac{p_{y}^{2}}{2M}-\frac{1}{2}M\omega'{}^{2}y^{2}\right)\psi_{k}(y)=\left(\frac{E^{2}}{2M}+\frac{1}{2}\omega-\frac{1}{8}M-\frac{Ma^{2}}{8}\right)\psi_{k}(y)\label{eq:79}
\end{equation}
with: 
\begin{equation}
y^{2}=\left(\xi+\frac{a}{4\omega'{}^{2}}\right)^{2}\label{eq:80}
\end{equation}
This leads to the equation of the harmonic oscillator: 
\begin{equation}
\left(p_{y}^{2}+\frac{1}{2}M\omega'{}^{2}y^{2}\right)\psi_{k}(y)=\bar{E}^{2}\psi_{k}(y)\label{eq:81}
\end{equation}
where: 
\begin{equation}
\bar{E}^{2}=\frac{E^{2}}{2M}-\frac{1}{2}\omega-\frac{1}{8}M-\frac{Ma^{2}}{8}\label{eq:82}
\end{equation}
The energy of the harmonic oscillator is: 
\begin{equation}
\overline{E}=\omega(n+\frac{1}{2})\label{eq:83}
\end{equation}
Thus, the final form of the energy levels is: 
\begin{equation}
E=\pm\sqrt{2M\omega(n+1)+\frac{M^{2}}{4}+\frac{M^{2}a^{2}}{16\omega^{2}-8a^{2}}}\label{eq:84}
\end{equation}
Equation (\ref{eq:84}) gives the energy spectrum of the Kemmer oscillator
in $(1+1)$-dimensional Rindler spacetime. The spectrum retains the
harmonic-oscillator structure but is modified by two relativistic
contributions: the rest-mass term $(M/2)$ and a correction proportional
to the acceleration parameter $a$. The appearance of the term $\tfrac{M^{2}a^{2}}{16(\omega^{2}-8a^{2})}$
indicates that acceleration effectively renormalizes the oscillator
frequency and removes the degeneracy of the energy levels. Physically,
the quantum number $n$ labels the oscillator excitations, while the
acceleration shifts their energies, with larger $a$ producing stronger
deviations from the flat-spacetime spectrum. In the limit $a\to0$,
Eq. (\ref{eq:84}) reduces to the well-known spectrum of the one-dimensional
Kemmer oscillator in Minkowski space, confirming the consistency of
the result.

It is important to note, however, that certain parameter ranges can
lead to unphysical results. In particular, in Eq. (\ref{eq:84}) the
denominator $16\omega^{2}-8a^{2}$ may vanish for specific choices
of the oscillator frequency $\omega$ and acceleration $a$. This
singularity has no physical meaning and instead signals the breakdown
of the approximations used in deriving Eq. (\ref{eq:84}). Similarly,
the radicand in Eq. (\ref{eq:84}) can become negative if the combination
of $M$, $\omega$, and $a$ falls outside the domain ensuring real
solutions, leading to imaginary (non-physical) energies. To guarantee
well-defined and real spectra, one must impose the restrictions
\begin{equation}
\omega^{2}>\tfrac{1}{2}a^{2},\qquad E^{2}\geq0,\label{eq:85}
\end{equation}
together with the condition that the denominator in Eq. (\ref{eq:85})
remain positive.

\begin{figure}
\begin{centering}
\includegraphics[scale=0.5]{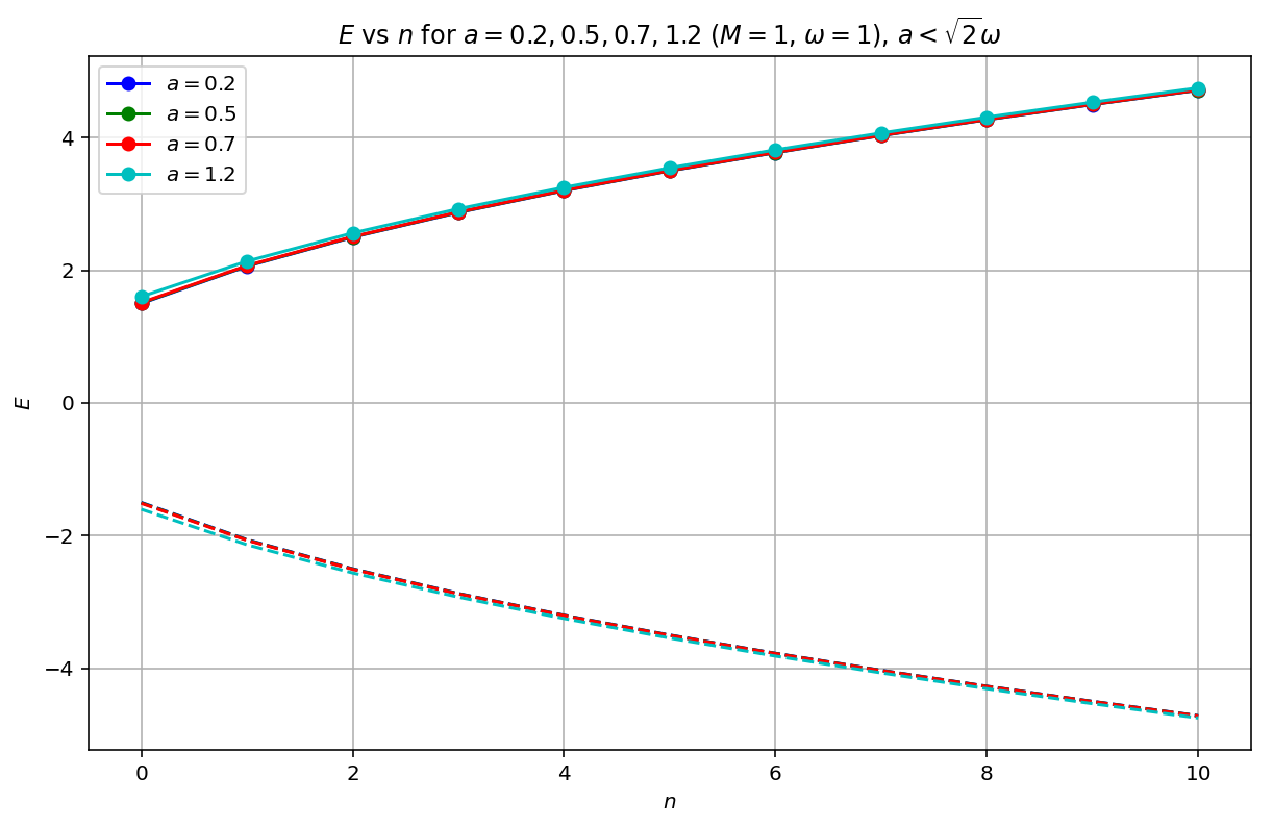}
\par\end{centering}
\caption{\label{fig:1}\foreignlanguage{american}{Energy Levels $E$ of the
one-dimensional Kemmer oscillator in Rindler spacetime as a function
of quantum number $n$ for various acceleration Parameters $a$ .}}
\end{figure}
\begin{figure}
\begin{centering}
\includegraphics[scale=0.5]{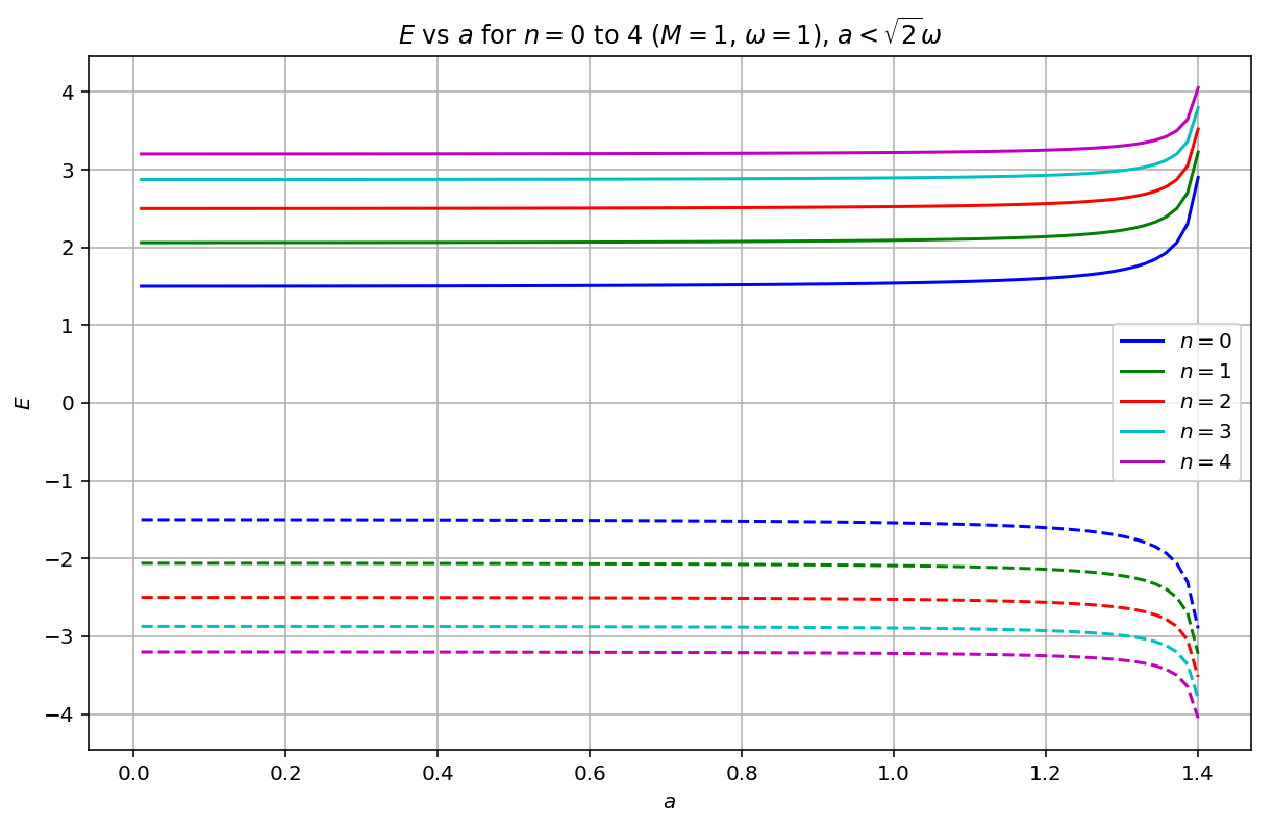}
\par\end{centering}
\caption{\label{fig:2}\foreignlanguage{american}{Energy spectrum $E$ as a
function of the acceleration parameter $a$ for the Kemmer oscillator
in 1+1 Rindler spacetime, with quantum numbers $\ensuremath{n=0},1,2,3,\ensuremath{4}$
($\ensuremath{a<\sqrt{2}\omega})$. Solid lines represent the positive
energy branch, while dashed lines indicate the negative branch. The
condition $a<\sqrt{2}\omega$ ensures the physical admissibility of
the spectrum, as derived from Eq. (92) in Section V, preventing singularities
in the denominator of the energy expression (Eq. (91)).}}
\end{figure}
\begin{figure}
\begin{centering}
\includegraphics[scale=0.5]{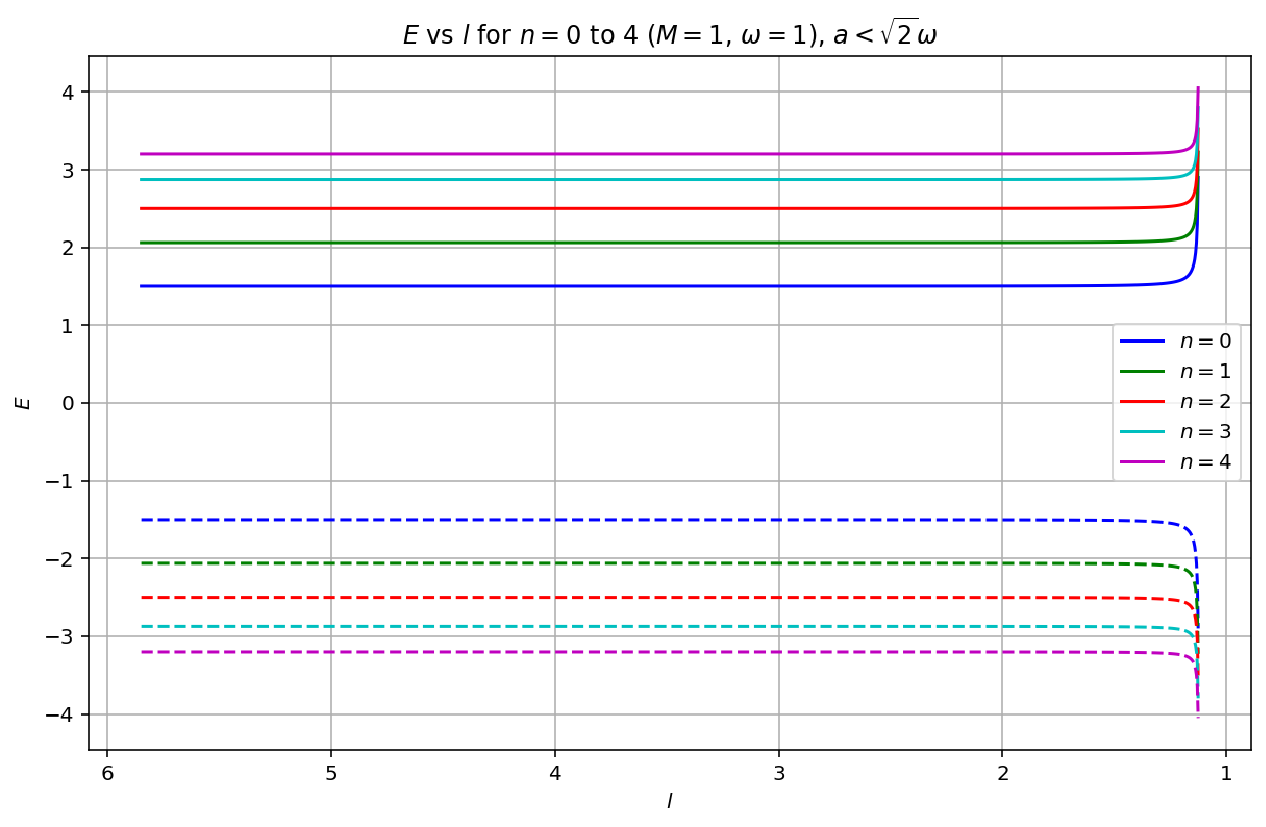}
\par\end{centering}
\caption{\label{fig:3}\foreignlanguage{american}{Energy spectrum $E$ as a
function of the characteristic length $l=(2/(M^{2}a))^{1/3}$ for
the Kemmer oscillator in 1+1 Rindler spacetime, with quantum numbers
$\ensuremath{n=0}to\ensuremath{4}$ ($a<\sqrt{2}\omega$). Solid lines
denote the positive energy branch, and dashed lines the negative branch.
The inverted x-axis accounts for the inverse relationship between
$l$ and $a$, reflecting the Rindler metric's spatial coordinate
transformation (Eq. (19), Section III). This visualization highlights
the acceleration-induced modification of the system's spatial scale,
a critical finding for understanding near-horizon kinematics and spectral
properties in curved spacetime.}}
\end{figure}

To illustrate and better interpret the final expression of the energy
spectrum, three figures have been generated.

In Figure. \ref{fig:1}, the dependence of the energy levels $E$
of the one-dimensional Kemmer oscillator in Rindler spacetime on the
quantum number $n$ is presented for various values of the acceleration
parameter $a$ ($a=0.02,0.05,0.7,1.2$, with $M=1$ and $\omega=1$).
The results illustrate the modifications to the spectrum arising from
uniform acceleration. Owing to the underlying curved Rindler metric
$ds^{2}=e^{2a\xi}(d\eta^{2}-d\xi^{2}),$ a gravitational-like influence
is introduced, altering the energy levels relative to those in flat
Minkowski spacetime. As the acceleration parameter increases, the
levels are shifted upwards, and the oscillator’s characteristic length
scale is deformed. These changes may lift degeneracies and modify
the level spacing. In the limiting case $a\to0$, the spectrum smoothly
reduces to the standard Kemmer oscillator spectrum in flat spacetime,
thus maintaining consistency with established results. 

In Figure. \ref{fig:2}, the energy spectrum of the free Kemmer field
in (1+1)-dimensional Rindler spacetime is displayed as a function
of the acceleration parameter $a$ for different quantum numbers $n$
($n=0,1,2,3,4$). This figure emphasizes the modifications to the
spectrum induced by the non-inertial frame. Uniform acceleration in
Rindler coordinates gives rise to an Unruh-like thermal effect with
effective temperature $T=a/(2\pi)$, which directly impacts the energy
levels. As $a$ increases, the levels experience an upward shift that
becomes more pronounced for higher quantum states. This trend reflects
the nontrivial coupling of acceleration to the spin-1 components of
the Kemmer field.

In Figure. \ref{fig:3}, the spectrum of the Kemmer oscillator including
a Dirac oscillator–type interaction in (1+1)-dimensional Rindler spacetime
is shown as a function of the characteristic length $l$ for quantum
states $n=0,1,2,3,4$. This analysis reveals the interplay between
the oscillator interaction and the geometric effects of Rindler spacetime.
The exponential factor of the metric, $e^{2a\xi}$, modifies the effective
potential and the spatial confinement properties, thereby deforming
the structure of the energy levels. The Dirac oscillator interaction,
expressed as a momentum shift $\vec{p}\to\vec{p}-iM\omega B\,\vec{r}$,
further enhances these modifications, producing shifts and broadening
of the energy levels with increasing acceleration. Moreover, this
interaction affects the normalizability of the wavefunctions, providing
deeper insights into the combined influence of acceleration-induced
curvature and oscillator dynamics in non-inertial frames. 

\section*{V. CONCLUSION}

\selectlanguage{american}%
In conclusion, we have derived exact solutions for the Kemmer oscillator
in (1+1)-dimensional Rindler spacetime, addressing the unique dynamics
of spin-1 bosons under uniform acceleration, including the Unruh effect
and their distinction from spin-0 and spin-1/2 systems. The introduction
of the Dirac oscillator interaction yields a closed-form spectrum,
with the acceleration parameter modifying the characteristic length,
shifting energy levels, and lifting degeneracies. Consistency with
the Minkowski limit ($a\to0$) validates the results against flat-spacetime
predictions. The clarified non-unitary transformation and verified
sign convention ensure mathematical rigor, while the defined parameter
domain ($\omega^{2}>\frac{a^{2}}{2}$) guarantees real energies. These
findings establish a tractable framework for exploring acceleration-induced
effects, with potential applications in quantum field theory, quantum
gravity, and analogue gravity platforms. Future work could extend
this analysis to higher dimensions .
\begin{acknowledgments}
The authors would like to express their sincere gratitude to the referee(s)
for their careful reading of the manuscript and for the insightful
comments and constructive suggestions provided. Their valuable feedback
has greatly contributed to improving the clarity, rigor, and overall
quality of this work.
\end{acknowledgments}

\selectlanguage{english}%

\section{Funding Statement}

No funds have been received for this manuscript.

\section{Data Availability Statement}

In this study, no new data was generated or analyzed.

\bibliographystyle{ChemEurJ}
\bibliography{/home/abdelmalek/Desktop/tarek/tarekUJK}

\end{document}